\documentstyle[11pt,paspconf,epsf,psfig]{article}

\begin{document}

\title{ChaMP and the High Redshift Quasars in X-rays}

\author{S. Mathur\altaffilmark{1}}
\affil{Harvard Smithsonian Center for Astrophysics
    Cambridge, MA 02138}
\author{H. Marshall\altaffilmark{1}}
\affil{MIT, Cambridge, MA 02139}
\author{N. Evans\altaffilmark{1}, P. Green\altaffilmark{1} and B.
Wilkes\altaffilmark{1} }
\affil{Harvard Smithsonian Center for Astrophysics
    Cambridge, MA 02138}

% Notice that some of these authors have alternate affiliations, which
% are identified by the \altaffilmark after each name.  The actual alternate
% affiliation information is typeset in footnotes at the bottom of the
% first page, and the text itself is specified in \altaffiltext commands.
% There is a separate \altaffiltext for each alternate affiliation
% indicated above.

\altaffiltext{1}{The Chandra Multiwavelength Project (ChaMP) is an independent
scientific collaboration for followup studies of serendipitous X-ray
sources in Chandra X-ray images.  The ChaMP Web site is
http://hea-www.harvard.edu/CHAMP.}
%\altaffiltext{2}{Society of Fellows, Harvard University}
%\altaffiltext{3}{Patron, Alonso's Bar and Grill}

% The abstract is entered in a LaTeX "environment", designated with paired
% \begin{abstract} -- \end{abstract} commands.  Other environments are
% identified by the name in the curly braces.

% Poster authors ONLY may omit the abstract in order to gain a little
% more page space for the text of the poster.

\begin{abstract}
{\it Chandra} X-ray Observatory, (formerly known as AXAF), will
 observe down to the
 flux limit of 2$\times 10^{-16}$ erg~s$^{-1}$~cm$^{-2}$. In its first
 year of operation {\it Chandra}'s CCD detectors will observe over 1500
  quasars serendipitously in the soft (0.5--3.5 keV) band.
 Over 200 quasars will be detected in X-rays in the redshift
 range $3<z<4$ and over 400 quasars in $2<z<3$. This will enable us to
 determine the high redshift X-ray luminosity function. This is the
 contribution by unabsorbed sources only. The total numbers would be
 larger by $\sim 60$\%.

\end{abstract}

% Keywords should be included, but they are not printed in the hardcopy.


% That's it for the front matter.  On to the main body of the paper.
% We'll only put in tutorial remarks at the beginning of each section
% so you can see entire sections together.

\section{Introduction}

NASA's {\it Chandra} X-ray Observatory was
launched on July 23, 1999. The {\it Chandra} Multiwavelength Project (ChaMP)
will combine radio to X-ray observations of serendipitous {\it
Chandra} sources, with emphasis on optical identification.
The ChaMP is superior to previous X-ray surveys because of (1)
unprecedented X-ray positional accuracy
 ($\sim 1^{\prime\prime}$), (2) X-ray flux limits 20 times deeper than current
 wide area surveys (down to $f(0.5-3.5 keV)\sim 2\times 10^{-16}$
erg~s$^{-1}$~cm$^{-2}$), (3) larger sky coverage ($\sim$ 8 deg$^2$)
per year than current deep surveys.
%For details of ChaMP see: http://hea-www.harvard.edu/CHAMP.

\section{Prediction of Redshift Distribution of Quasars in ChaMP Fields}

{\bf The X-ray Luminosity Function} at $z=0$ is described as

$$\Phi(L_X)=\Phi^{\star}_1 L^{-\gamma1}_{44} ~~~~for ~L<L^{\star}(0)$$

$$\Phi(L_X)=\Phi^{\star}_2 L^{-\gamma2}_{44} ~~~~for
{}~L>L^{\star}(0)$$\\

\noindent
where L$_{44}$ is the X-ray luminosity in $10^{44}$
erg~s$^{-1}$. The redshift evolution of the luminosity
function is characterized by

$$ L_{X}(z)= L_{X}(0) (1+z)^k $$

Continuity of the luminosity function at the break luminosity requires
that

$$\Phi^{\star}_1=\Phi^{\star}_2 L^{\star (\gamma1-\gamma2)}_{44} $$

The total number $N$ of quasars in the sample is obtained by
integrating the luminosity function over luminosity and volume, i.e.,

$$ N=\int\!\int \Phi(L_X, z)\Omega(L_X, z) dV(z) dL_X $$

Here $\Omega(L_X, z)$ is the solid angle covered by the survey as a
function of redshift and luminosity. The parameters of the X-ray
luminosity function determined by Boyle et al. (1993) are as
follows: $\gamma1=1.7 \pm 0.2$, $\gamma2=3.4 \pm 0.1 $, $\log
L^{\star}(0)=43.84$,  $\Phi^{\star}_1=5.7 \times 10^{-7} Mpc^{-3}
(10^{44} erg~ s^{-1})^{\gamma1-1}$. Following Comastri et al. (1995),
we have used k=2.6 and increased the the normalization
$\Phi^{\star}_1$ by 20\%.

{\bf The X-ray logN-logS Curve}: Using the above luminosity function
we derived the number density of
quasars as a function of observed flux. The
luminosity function was
integrated over the luminosity range $10^{42}<L_X<10^{48}$ erg s$^{-1}$
and the redshift range $0<z<4$. H$_0=50$ and q$_0=0$ were assumed
throughout. The predicted logN-logS curve is shown in
 figure 1.

\begin{figure}[h]
%\epsscale{0.80}
\psfig{figure=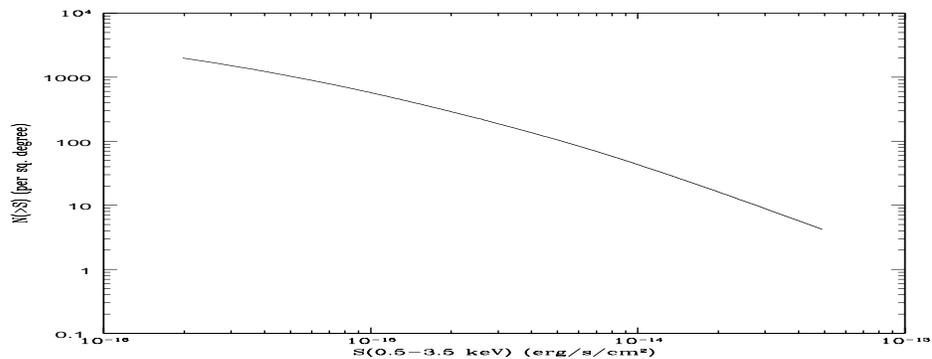,height=2in,width=5in}
\caption{The predicted number counts in the soft band for unabsorbed
quasars.}
%\caption{The total sky coverage of ChaMP fields as a
%function of flux limit in the soft band.}
\end{figure}

Since the unabsorbed sources dominate at the faint end in the soft X-ray
range, and since they are
likely to be observed at high redshift, in the present analysis we
will concentrate on unabsorbed sources only.  The absorbed sources
would contribute an additional $\sim$ 60\% (Comastri et al. 1995),
making the total number
consistent with the extrapolation of the empirical determination of logN-logS
(Hasinger
et al. 1993). The flux of unabsorbed
quasars is given by $f \propto E^{-\alpha}$ and in the soft X-ray band,
$\alpha$ is typically 1.3.

{\bf The ChaMP Sky Coverage}:  The ChaMP Cycle 1
consists of 85 extragalactic fields, |b|$>20^{o}$. From all the Chandra
cycle 1 fields we have excluded (1) deep fields of PI survey
observations, (2) fields with extended sources \& planetary targets,
(3) ACIS sub-arrays and continuous clocking modes. See figure 2 for
ChaMP sky coverage as a function of flux limit.

\begin{figure}[h]
%\epsscale{0.80}
\psfig{figure=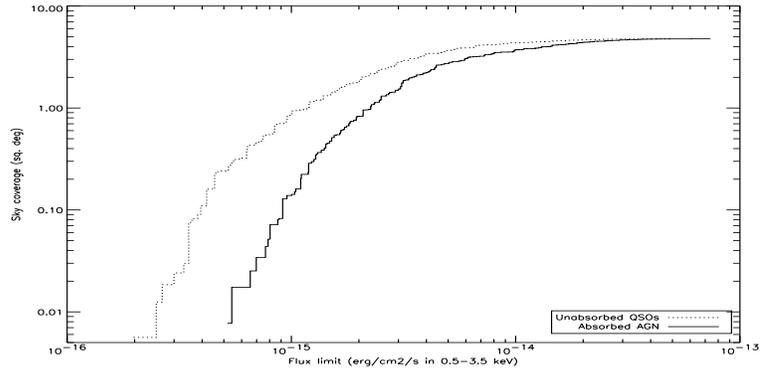,height=2in,width=4in,angle=90}
%\vspace*{-1.5in}
\caption{The total sky coverage of ChaMP fields as a
function of flux limit in the soft band.}
\end{figure}

{\bf Cumulative Number Distribution in ChaMP:} Integrating the
predicted logN-logS over the ChaMP sky coverage, we
obtained the cumulative number distribution of quasars in the ChaMP
fields (figure 3). The total number in  soft band is
 expected to be over 1500 for unabsorbed sources and over 2500 total.

\begin{figure}[h]
%\epsscale{0.80}
\psfig{figure=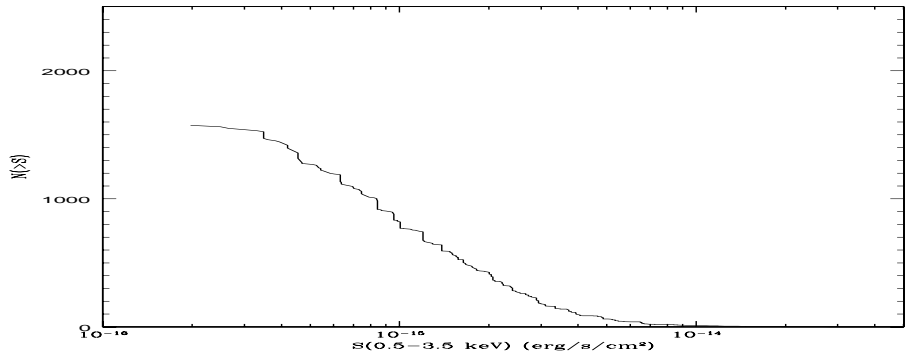,height=2in,width=5in}
\caption{Expected cumulative source counts. Unabsorbed sources only.}
\end{figure}

{\bf Predicted Redshift Distribution:} The histogram (figure 4) shows the
predicted number distribution of quasars
in ChaMP fields. Over 200 quasars will be detected in the redshift
range $3<z<4$ and over 400 quasars in $2<z<3$.

\begin{figure}[h]
%\epsscale{0.80}
\psfig{figure=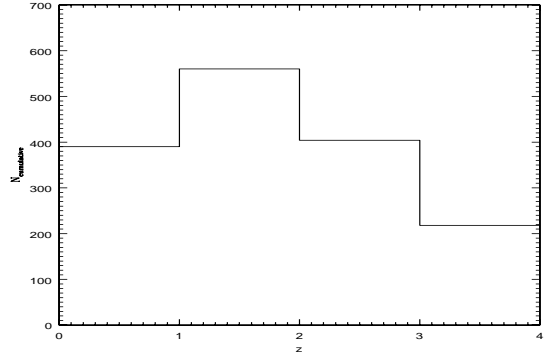,height=2in,width=3in}
\caption{Redshift distribution of the unabsorbed sources expected to
be detected in ChaMP fields.}
\end{figure}

%\newpage
\section{Comparison with Previous X-ray Surveys}

\begin{table}[h]
\begin{tabular}{|lrr|}
\hline\hline
Survey & Total number of Sources & Quasars at $z>2$ \\
\hline
EMSS & 835 & $<5$ \\
(Gioia et al.)& & \\
ROSAT Deep & 661 & 12 \\
(Hasinger et al.)& & \\
ROSAT & 89 & $<10$ \\
(Boyle et al.)& & \\
ChaMP & $>1500$ & $>600$\\
(soft band, unabsorbed)& & \\
\hline
\hline
\end{tabular}
\end{table}

We will be able to determine the X-ray luminosity function and its
redshift evolution with unprecedented accuracy.\\

% Now comes the reference list.  Since we typed out the citations ourselves,
% the reference list is enclosed in a "references" environment.  Each
% new reference begins with a \reference command which sets up the proper
% indentation.  Typography that may be required in the reference list by
% the editorial staff must be included by the author.
%
% Observe the "standard" order for bibliographic material: author name(s),
% publication year, journal name, volume, and page number for articles.
% Some journal names are available as macros; see the WGAS markup
% instructions for a listing of which ones have been "macro-ized".
% Note the use of curly braces to delimit the font changes: it is essential
% that this be done to limit the scope of the font declaration.
%
% There is no need to engage in any other typographic manipulation.

\vspace{0.2in}
It's my pleasure (SM) to thank A. Comastri for useful discussions. This
work is supported in parts by NASA grant NAG5-3249 (LTSA).

% That's all, folks.
%
% The technique of segregating major semantic components of the document
% within "environments" is a very good one, but you as an author have to
% come up with a way of making sure each \begin{whatzit} has a corresponding
% \end{whatzit}.  If you miss one, LaTeX will probably complain a great
% deal during the composition of the document.  Occasionally, you get away
% with it right up to the \end{document}, in which case, you will see
% "\begin{whatzit} ended by \end{document}".

\end{document}